\documentclass[12pt]{article}
\usepackage{amssymb,epsf}
\def\lsim{\mathrel{\rlap {\raise.5ex\hbox{$ < $}}
{\lower.5ex\hbox{$\sim$}}}}
\def\gsim{\mathrel{\rlap {\raise.5ex\hbox{$ > $}}
{\lower.5ex\hbox{$\sim$}}}}
\def\sqr#1#2{{\vcenter{\vbox{\hrule height.#2pt
        \hbox{\vrule width.#2pt height#1pt \kern#1pt
           \vrule width.#2pt}
        \hrule height.#2pt}}}}

\def\lsim{{\displaystyle
{{\raise-8pt\hbox{$ <$}}
\atop{\raise5pt\hbox{$\sim$}}}}}
\def\gsim{{\displaystyle
{{\raise-8pt\hbox{$ >$}}
\atop{\raise5pt\hbox{$\sim$}}}}}
\def\slsim{{\displaystyle
{{\raise-8pt\hbox{$\scriptstyle <$}}
\atop{\raise5pt\hbox{$\scriptstyle \sim$}}}}}
\def\sgsim{{\displaystyle
{{\raise-8pt\hbox{$\scriptstyle  >$}}
\atop{\raise5pt\hbox{$\scriptstyle \sim$}}}}}
\newskip\humongous \humongous=0pt plus 1000pt minus 1000pt

\newcommand{\sumpf}[0]{\sum_{(H^{\rm f},G^{\rm f})}\! \! \! \!
{\raise
4pt
\hbox{$'$}}\,}

\newcommand{\sump}[0]{\sum_{(h,g)}\!{\raise 4pt \hbox{$'$}}\,}

\def\bs{\begin{subequations}}
\def\es{\end{subequations}}
\catcode`\@=11
\newcount\hour
\newcount\minute
\newtoks\amorpm
\hour=\time\divide\hour by60
\minute=\time{\multiply\hour by60 \global\advance\minute by-\hour}
\edef\standardtime{{\ifnum\hour<12 \global\amorpm={am}%
        \else\global\amorpm={pm}\advance\hour by-12 \fi
        \ifnum\hour=0 \hour=12 \fi
        \number\hour:\ifnum\minute<10 0\fi\number\minute\the\amorpm}}
\edef\militarytime{\number\hour:\ifnum\minute<10 0\fi\number\minute}
\def\draftlabel#1{{\@bsphack\if@filesw {\let\thepage\relax
   \xdef\@gtempa{\write\@auxout{\string
      \newlabel{#1}{{\@currentlabel}{\thepage}}}}}\@gtempa
   \if@nobreak \ifvmode\nobreak\fi\fi\fi\@esphack}
        \gdef\@eqnlabel{#1}}
\def\@eqnlabel{}
\def\@vacuum{}
\def\draftmarginnote#1{\marginpar{\raggedright\scriptsize\tt#1}}
\def\draft{\oddsidemargin -.2truein
        \def\@oddfoot{\sl preliminary draft \hfil
        \rm\thepage\hfil\sl\today\quad\militarytime}
        \let\@evenfoot\@oddfoot \overfullrule 3pt
        \let\label=\draftlabel
        \let\marginnote=\draftmarginnote
   \def\@eqnnum{(\theequation)\rlap{\kern\marginparsep\tt\@eqnlabel}%
\global\let\@eqnlabel\@vacuum}  }
\def\subequations{\refstepcounter{equation}%
  \edef\@savedequation{\the\c@equation}%
  \@stequation=\expandafter{\theequation}
  \edef\@savedtheequation{\the\@stequation}
  \edef\oldtheequation{\theequation}%
  \setcounter{equation}{0}%
  \def\theequation{\oldtheequation\alph{equation}}}

\def\endsubequations{\setcounter{equation}{\@savedequation}%
  \@stequation=\expandafter{\@savedtheequation}%
  \edef\theequation{\the\@stequation}\global\@ignoretrue
  \vspace*{-12pt} \\}
\def\bs{\begin{subequations}}
\def\es{\end{subequations}}
\relax
\def\Im{\,{\rm Im}\, }

\def\thefootnote{\fnsymbol{footnote}}
\def\be{\begin{equation}}
\def\ee{\end{equation}}
\def\ba{\begin{eqnarray}}
\def\ea{\end{eqnarray}}

\def\ee{\end{equation}}
\def\bea{\begin{eqnarray}}
\def\eea{\end{eqnarray}}
\def\nn{\nonumber}

\def\np#1#2#3{Nucl. Phys. {\bf{B#1}} (#2) #3}
\def\pl#1#2#3{Phys. Lett. {\bf{B#1}} (#2) #3}

\def\pr#1#2#3{Phys. Rev. {\bf{D#1}} (#2) #3}

\newcommand{\uarrw}[0]{\mathrel{
{\raise.5ex\vbox{\hrule width 1cm}\hskip-6pt\rightarrow}}}
\def\thebibliography#1{%
\vskip 0.5cm \centerline{\bf References}
\list{%
[\arabic{enumi}]}{\settowidth\labelwidth{[#1]}
\leftmargin\labelwidth
\advance\leftmargin\labelsep
\usecounter{enumi}}
\def\newblock{\hskip .11em plus .33em minus .07em}
\sloppy\clubpenalty4000\widowpenalty4000
\sfcode`\.=1000\relax}

\renewcommand{\theequation}{\arabic{section}.\arabic{equation}}

\renewcommand{\section}{\setcounter{equation}{0}\@startsection%
{section}{1}{0mm}{-\baselineskip}{0.5\baselineskip}%
{\normalfont\normalsize\bfseries}}

\renewcommand{\subsection}{\@startsection%
{subsection}{2}{0mm}{-\baselineskip}{0.5\baselineskip}%
{\normalfont\normalsize\slshape}}
\begin{document}
\renewcommand{\theequation}{\arabic{section}.\arabic{equation}}
\begin{titlepage}
\begin{flushright}
Bicocca-FT/00/03,\\
hep-th/0002214 
\end{flushright}
\begin{centering}
\vspace{.15in}
{\bf \large
Non-perturbative string connections}$^\ast$
\\
\vspace{1.5 cm}
{Andrea Gregori}$^{\ 1}$\\
\medskip
\vspace{.4in}
{\it  Universit{\`a} di Milano--Bicocca},
{\it  via Celoria 16, 20133  Milano}\\
\vspace{2.4cm}
{\bf Abstract}\\
\vspace{.1in}
We discuss the  duality between two type I compactifications 
to four dimensions and an heterotic construction 
with spontaneous breaking of the ${\cal N}=4$ supersymmetry to
${\cal N}=2$.
This duality allows us to gain insight into the non-perturbative properties
of these models. Through the analysis of the gravitational corrections,
we then investigate the connections between four-dimensional, $N=2$
M-theory vacua constructed as orbifolds of type II, heterotic, and type I
strings. 
\end{centering}
\vspace{2,4cm}
\hrule width 6.7cm
$^\ast$\  Talk given at the TMR meeting on {\sl Quantum aspects of gauge 
theories, supersymmetry and unification}, Paris,
1--7 September 1999.\\
\\
$^1$e-mail: agregori@pcteor.mi.infn.it

\end{titlepage}
\newpage
\setcounter{footnote}{0}
\renewcommand{\thefootnote}{\arabic{footnote}}

\setcounter{section}{1}

After the ``second string revolution'', that took place after 
1995, it is a common belief that all the string constructions
are manifestations of a unique underlying theory.
In most of the cases, the different string models correspond
to different regions in the moduli space of the underlying theory.
There are however several cases in which apparently disconnected
vacua turn out to be indeed equivalent, being simply related by a
``change of parametrization''.

In some cases, this reparametrization maps perturbative moduli
into non-perturbative ones. In these cases, knowing this map
allows to compute in an easy way quantities that would be beyond
a perturbative, often short, computation.
This is the case for instance of the duality between 
the heterotic string compactified to four dimensions and
the type IIA string compactified on a K3 fibration \cite{ht}.
The heterotic dilaton--axion field is mapped into a perturbative
modulus, associated to the volume form of the base of the fibration, on the
type IIA side. Therefore, what is non-perturbative on the heterotic side 
is perturbative on the type IIA side.
This relation, that recently received a further confirmation \cite{kp}, 
has been used in order to compute  
the non-perturbative correction to the effective coupling of the 
$R^2$ term in some specific examples \cite{hm}--\cite{gkp2}
(An analogous relation exists also between some type II asymmetric
orbifold compactifications and the type IIA string \cite{gkp2}--\cite{gkr}).
However, the cases in which such duality exists are far from covering 
the main part of the heterotic constructions. It seems indeed that
some of the most interesting cases don't fall in this class.
The question is therefore whether it is nevertheless possible,
in some of these cases,
to obtain (at least partial) information on the non-perturbative behavior.
We present here an analysis of a four-dimensional,
${\cal N}=2$ heterotic construction
which, although without type IIA dual, possesses nevertheless a
pair of type I string duals. Putting together the informations
coming from all these constructions, it is possible to obtain
partial but not irrelevant informations about the non-perturbative
behavior of this model.

Through the analysis of the gravitational corrections,
we discuss then the connections of these constructions
with other type II and heterotic constructions. 

\vspace{.4cm}
\begin{centering}
$* ~ * ~ *$\\
\end{centering}
\vspace{.5cm}

The heterotic model we consider is constructed as
a $Z_2$, ``freely acting'' orbifold of the string compactified on 
$T^6=T^2 \times T^4$. The $Z_2$ projection acts as a reflection,
$x_i \to - x_i$, on $T^4$ and as a half-circumference translation 
in one circle of $T^2$. 
Imposing modular invariance and requiring
the shift of the momenta of the $T^2$-lattice
produced by this translation to be left-right symmetric, leads to
a specific embedding of the spin connection into the gauge group,
such that, at the orbifold point, the massless spectrum
originating from the $c=(0,16)$ currents always contains
an equal number of vector and hypermultiplets.
This equality is however not due to a ``level two'' realization of
the gauge group: the rank is sixteen, i.e. the maximal
allowed in perturbative heterotic constructions.
The free action of $Z_2$ produces then a spontaneous breaking of the
${\cal N}=4$ supersymmetry, that can be restored when the 
translated coordinated is decompactified. This corresponds to a
special limit in the space of the moduli of the two-torus,
$T$ and $U$, associated respectively to the K\"{a}hler class and the complex 
structure. At the $U(1)^{16}$ point, this model has
sixteen vector multiplets and sixteen hypermultiplets from the currents, 
plus the three vectors and four hypermultiplets from the compact space.
Due to the free action of $Z_2$, there are no massless states
originating from the twisted sector.
Although it exists a type IIA orbifold construction with the same massless
spectrum, obtained by compactification on an orbifold limit of a
Calabi Yau manifold with Hodge numbers (19,19), 
this is not dual to the heterotic orbifold;
the CY$^{19,19}$ manifold is not in fact a K3 fibration.

The heterotic orbifold
possesses on the other hand a type I dual, constructed as an orientifold 
of the ${\cal N}=4$, type IIB string, in which the ${\cal N}=8$
supersymmetry is spontaneously broken by a $Z_2$ freely acting projection,
whose action of $T^6$ is the same as that of the heterotic model.  
In the case of the heterotic string, T-duality makes irrelevant
the choice of the translation, that can equivalently be performed 
on the momenta, projection $(-1)^m$, or on the windings,
projection  $(-1)^n$;
in the case of type I orientifolds on the other hand
the two projections lead to rather different models \cite{adds}.
In the first case, the model obtained is
exactly equivalent to the heterotic one, with a perturbative,
spontaneous breaking of the ${\cal N}=4$ supersymmetry
and a gauge group of rank sixteen, originating entirely 
from D9-branes: there are no D5-branes. By introducing appropriate Wilson
lines it is possible to obtain exactly the same spectrum of the
heterotic construction.
The type I orientifold obtained from the type IIB string with
translation $(-1)^n$, on the other hand, has still a gauge group
of rank sixteen, but that now originates half from the D9-branes sector
and half from the D5-branes sector. Moreover, the spectrum on the branes 
doesn't feel the breaking of supersymmetry.
This model doesn't look therefore dual to the previous constructions.

In order to investigate the conjectured duality between the heterotic
and the first of the two type I constructions, 
we consider the string corrections
to the effective coupling of the $R^2$ term.
As it was discussed in Refs. \cite{gkp,gkp2,gk},
in order to compare string constructions, it is
necessary to project out the non-universal contribution
coming from the coupling of the ``bulk'' sector with the currents. 
This can be consistently done because the contribution of the ``currents''
to the gravitational and gauge couplings are proportional, and
can be subtracted by properly redefining the amplitude we compute. 
On the heterotic side, a further contribution must be subtracted,
namely that coming at the singularities in the space of the moduli
$T$ and $U$: at these points, new vector and/or hypermultiplets
appear. This phenomenon doesn't happen on the type I side.
Once this projection has been performed, the renormalization 
of the gravitational amplitudes read, in the two models:
\ba
{\rm Het.}:~~~~~ 
{16 \, \pi^2 \over g^2} & = &
16 \, \pi^2 \, \Im S -2 \log \Im T \vert \vartheta_4 ( T) \vert^4 \nn \\
&& -2 \log \Im U \vert \vartheta_4 ( U) \vert^4 \,
+ \, {\cal O}(\log \mu \big/ M) \, , \label{het}\\
&& \nonumber \\
{\rm Type\, I}:~~~~~
{16 \, \pi^2 \over g^2} & = &
16 \, \pi^2 \, \Im S 
-2 \log \Im U \vert \vartheta_4 ( U) \vert^4 \nonumber \\
&& + {\cal O}(\log \mu \big/ M)\, .
\label{type1}
\ea
Apparently, a term is missing in Eq. (\ref{type1}).
The solution to this puzzle comes from the observation that the second term
in the r.h.s. of Eq.(\ref{het}) behaves, for large-$T$, as:
$\log \Im T \vert \vartheta_4 ( T) \vert^4 \sim \log \Im T$.
There is no linear divergence in $T$, and the logarithmic behavior can
be interpreted as due to non-perturbative phenomena. Indeed, it can
be removed by switching on an appropriate infrared cut-off (see Ref. 
\cite{solving}).
The corrections (\ref{het}) and (\ref{type1}) are therefore consistent with
the duality of these constructions. We may however ask what is indeed
the meaning of the heterotic field $T$, whose contribution is apparently
lost when going to the type I dual.
In the opposite limit, $T \to 0$, the heterotic correction (\ref{het})
diverges linearly in the inverse field, $\tilde{T} \equiv -1 \big/ T$.
By redefining $\tilde{T} \equiv S^{\prime}$, we see that
in the large $\tilde{T}$ limit, (\ref{het}) behaves as:
\be
{16 \, \pi^2 \over g^2} \, \approx \,  
16 \, \pi^2 \, \Im S \, + \, 16 \, \pi^2 \, \Im S^{\prime} 
-2 \log \Im U \vert \vartheta_4 ( U) \vert^4 \, .
\ee   
This is indeed the behavior of a type I model with 
both D9- and D5-branes sectors.
Actually, this is the correction as it would be computed on the
second type I construction.
We guess that indeed the second type I construction \emph{is dual}
to the heterotic and the other type I model.
This hypothesis is supported by a look at the path followed
by the D5-branes when their coupling, parametrized by $S^{\prime}$,
becomes very weak ($S^{\prime} \to 0$) \cite{gk}.
The crucial point is that this model can be viewed as obtained via
a freely acting $Z_2$ projection performed along the eleventh coordinate of 
the M-theory \cite{adds}. 
In this model, owing to the free action of this projection,
S(S$^{\prime}$)-duality is broken. Indeed, in the limit
$S^{\prime} \to 0$, that corresponds to
a decompactification of the $T^4$, the D5-branes 
massless fields are still present in the spectrum. However, being the D5-branes
coupling missing, these states have to be interpreted as originating from
D9-branes. 
Supersymmetry of the branes spectrum is indeed an artifact, due
to the separation into closed and open sectors, typical of type I 
constructions. It seems therefore possible to imagine to break
supersymmetry only on the bulk, represented by the closed string sector.
However, through interactions of the bulk fields with 
the fields living on the branes, the breaking of supersymmetry
is then communicated to the whole theory.
It is indeed from the heterotic dual, where there is no
such a fake separation, that we learn how the supersymmetry breaking is 
actually communicated from one sector to the other.

The relation to the M-theory is better understood by going to the dual,
type I$^{\prime}$ picture, where D9- and D5-branes appear as
D4- and D8- branes, and the eleventh coordinate corresponds to the tenth
coordinate of the string.
From the M-theory point of view, the freely acting, Scherk--Schwarz 
supersymmetry breaking projection
on the eleventh coordinate leaves untouched, at least in a first approximation,
the two ``Ho\v{r}ava--Witten walls'', whose massless fields
correspond respectively to the D4- and D8-branes fields \cite{adds}.
Since the system is symmetric under reflection of the eleventh 
coordinate, we argue that  the breaking of the S$^{\prime}$-duality
on the D4 sector has a mirror in an analogous breaking of S-duality
on the D8-branes sector.
Under this hypothesis, the correction given in Eq. (\ref{het})
should be promoted to:
\begin{eqnarray}
{16 \, \pi^2 \over g^2} & = &
-2 \log \Im \tilde{S} \vert \vartheta_4 ( \tilde{S} ) \vert^4
-2 \log \Im \tilde{S}^{\prime} \vert \vartheta_4 ( \tilde{S}^{\prime} ) \vert^4
\nonumber \\
&& -2 \log \Im U \vert \vartheta_4 ( U ) \vert^4 
~+ {\cal O}\left( {\rm e}^{-(\tilde{S},\tilde{S}^{\prime},U)} \right) \, ,
\nonumber \\
&&
\label{gnp}
\end{eqnarray}
where $\tilde{S} \equiv -1 \big/ S$, 
$\tilde{S}^{\prime} \equiv -1 \big/ S^{\prime}$, and the last
term is a series of exponentials, symmetric in the three fields,
suppressed in the large and small fields limits.
This is the typical behavior of a theory with spontaneous breaking
of the ${\cal N}=8$ supersymmetry.
Expression (\ref{gnp}) diverges linearly for large-$S$,
large-$S^{\prime}$ (and small $U$), but only logarithmically
for large $\tilde{S}$, $S^{\prime}$, $U$. This is the limit of restoration of 
the ${\cal N}=8$ supersymmetry.

\vspace{.3cm}
\begin{centering}
$* ~ * ~ *$\\
\end{centering}
\vspace{.4cm}

The analogous gravitational correction in the type IIA
orbifold with Hodge numbers (19,19), is (see Ref. \cite{gkr}):
\be
{16 \, \pi^2 \over g^2}  =  
-2 \log \Im T^{(1)} \vert \vartheta_4 ( T^{(1)} ) \vert^4 
-6 \log \Im T^{(2)} \vert \eta ( T^{(2)} ) \vert^4 
-6 \log \Im T^{(3)} \vert \eta ( T^{(3)} ) \vert^4 \, , 
\label{1616}
\ee
where $T^{(1)}$, $T^{(2)}$, $T^{(3)}$ are the moduli associated to the
K\"{a}hler classes of the three tori of $T^6=T^2 \times T^2 \times T^2$,
and we omit for simplicity the term logarithmically dependent on 
the infrared cut-off.
It is clear that this correction doesn't match the heterotic one.
Indeed, this type IIA orbifold is dual to a type II asymmetric orbifold,
in which the modulus $T^{(1)}$ plays the role of the dilaton--axion field.
It is also clear that, if the heterotic model has a spontaneously broken
${\cal N}=8$ supersymmetry, it cannot possess a type IIA dual, in which
this super-Higgs phenomenon would appear as perturbative.
Indeed, a type IIA orbifold with the spontaneous breaking of ${\cal N}=8$
exists: it is a $Z_2 \times Z_2$ orbifold in which both the 
projections act freely \cite{sv,gkr}. 
All the states of the twisted sectors are massive,
and the massless spectrum contains only, besides the gravity multiplet,
three vector multiplets and four hypermultiplets.
The gravitational correction reads:
\be
{16 \, \pi^2 \over g^2}  =  
-2 \log \Im T^{(1)} \vert \vartheta_4 ( T^{(1)} ) \vert^4 
-2 \log \Im T^{(2)} \vert \vartheta_4 ( T^{(2)} ) \vert^4 
-2 \log \Im T^{(3)} \vert \vartheta_4 ( T^{(3)} ) \vert^4 \, . 
\label{00}
\ee
In a particular limit in the moduli space ($T^{(1)} \to 0$), 
this orbifold behaves as
a K3 fibration, and can be compared to an heterotic dual with
the same massless spectrum \cite{gkp2}. 
The map of moduli is in that case: $T^{(1)} \to -1 \big/ S$,
$T^{(2)} \to T$, $T^{(3)} \to U$.
In the opposite limit, $T^{(1)} \to \infty$, the model matches
instead a type II asymmetric, freely acting orbifold, with map of moduli:
$T^{(1)} \to S$, $T^{(2)} \to T$, $T^{(3)} \to U$. Heterotic and type II 
asymmetric orbifolds turn out therefore to be S-dual the one to the other.

Although not coinciding in the ``bulk'' of the moduli space, 
the corrections (\ref{gnp}) and (\ref{1616}) match at the
corner of moduli space, precisely at the limit in which
the supersymmetry breaking $Z_2$ projection of the heterotic/type I side
ends to act freely: in this limit, corresponding to
$S^{\prime} \to \infty$, $U \to 0$, (\ref{gnp}) matches (\ref{1616})
in the limit $T^{(2)} \to \infty$, $T^{(3)} \to \infty$.
This is however not evidence that at the limit the two theories really match:
indeed, when we recover a genuine orbifold limit, 
we expect new massless states to appear,
associated to the orbifold fixed points\footnote{In the light
of recent investigations, this part of 
the conclusions presented in
\cite{gk} needs a slight modification.}.
On the other hand, it is possible to interpolate between the
freely acting type IIA/heterotic/type II asymmetric orbifolds and
the heterotic/type I constructions with rank 16 by switching on
an appropriate Wilson line that, already at the ${\cal N}=4$ level,
lifts the mass of all the fields originating from the currents.
It turns out therefore that some of these constructions can be connected
by passing through appropriate limits in the moduli spaces.

\vskip 1.cm
\centerline{\bf Acknowledgements}
\noindent
This talk is based on a work done in collaboration with C. Kounnas;
I take this opportunity to address my thanks to him, and
to acknowledge the EEC, under the contract
TMR-ERBFMRX-CT96-0045, the Swiss National Science Foundation
and the Swiss Office for Education and Science, for financial support.

\end{document}